# Theory of small amplitude bimodal atomic force microscopy in ambient conditions

Sergio Santos

Departament de Disseny i Programació de Sistemes Electrònics**,** UPC - Universitat Politècnica de Catalunya Av. Bases, 61, 08242 Manresa (Barcelona), Spain

Abstract

Small oscillation amplitudes in dynamic atomic force microscopy can lead to minimal invasiveness and high resolution imaging. Here we discuss small oscillation amplitude imaging in the context of ambient conditions and simultaneously excite the second flexural mode to access contrast channels sensitive to variations in sample's properties. Two physically distinct regimes of operation are discussed, one where the tip oscillates above the hydration layer and another where the tip oscillates in perpetual contact with it. It is shown that the user can control the region to be probed via standard operational parameters. The fundamental theory controlling the sensitivity of the second mode phase shift to compositional variations is then developed. The second mode phase shift is controlled by an interplay between conservative tip-sample interactions, energy transfer between modes and irreversible loss of energy in the tip sample junction.

Keywords: bimodal, small amplitude, dissipation, phase shift, energy transfer

I. Introduction

In recent years there has been increasing interest in small amplitude imaging in dynamic atomic force microscopy (AFM) because it can lead to high resolution and minimally invasive mapping as the amplitudes become comparable to intermolecular bonds[1-5]. This interest in small amplitudes has been a continuous trend in the three main working environments, namely liquid[2, 4, 6, 7], vacuum[1, 8, 9] and ambient[3, 10, 11]. In ambient conditions, the nanometer thick water film covering surfaces and the capillary interactions that result when the sharp tip of the AFM comes close to the surface complicate the theory[12-14], the interpretation of data[3, 15] and experimentation[3, 16, 17]. It has also recently been shown that reaching the localized short-range surface forces that exist under the hydration layer [3] might be a prerequisite to achieving atomic resolution and resolving[11] the double helix of single DNA molecules. Increasing lateral resolution while reducing peak forces and invasiveness is not sufficient in terms of what the community requires from dynamic AFM[18-20]. In particular, there is increasing interest in simultaneously mapping sample composition while increasing sensitivity to variations in sample's properties [19, 21-23] of systems presenting nanoscale heterogeneity [24, 25]. In summary, the theory and experimental realization of the potential of small amplitudes and high resolution however is still emerging[3, 7, 11].

Here the theory of small amplitude dynamic atomic force microscopy is discussed and developed in the context of ambient conditions and with an emphasis on amplitude modulation (AM) AFM. Small amplitudes are defined as those in the nm or sub-nano-meter range throughout since these are comparable to intermolecular bonds or small molecules.

Two experimentally accessible and distinct regimes of operation, namely the non-contact NC and the small amplitude small set-point SASS regions or regimes, are discussed. The prediction is that standard operational parameters in commercial AFMs can be employed to reach the full range of distances of interest for high resolution and minimally invasive operation. Then the second mode is externally excited with sub-angstrom amplitudes to make contrast channels sensitive to compositional contrast[21] experimentally accessible[22, 26]. The sensitivity of the phase shifts to conservative and dissipative interactions is quantified. The discussion of energy dissipation is limited to tens of meV but shows that second mode phase shifts in the order of one degree or more follow. It is shown that both conservative interactions and energy transfer between modes are responsible for phase contrast, i.e. first and second modes, when conservative forces only are present in the interaction as it occurs in standard monomodal AFM in liquid[27]. When dissipation is allowed, phase contrast originates from a combination between conservative interactions, energy transfer between modes and irreversible loss of stored energy per cycle.

## II. Tip position and small amplitude imaging

The dynamics of a cantilever in dynamic AFM can be approximated by M modal equations of motion coupled via the non-linear, and typically unknown, tip-sample force $F_{ts}$[26]

$$\frac{k_m}{\omega_{0m}^2}\ddot{z}_m(t) + \frac{k_m}{Q_m\omega_{0m}}\dot{z}_m(t) + k_m z_m = F_{01}\cos(\omega_1 t) + F_{02}\cos(\omega_2 t) + F_{ts}$$

(1)

In (1) m stands for mode, $k_m$, $\omega_{0m}$ and $Q_m$ are the modal stiffness, resonance frequency and Q factor of the m mode, $F_{01}$ and $F_{02}$ are the magnitudes of two external driving forces at frequencies $\omega_1$ and $\omega_2$ respectively and $z_m$ is the deflection of the cantilever for mode m. The number of external driving forces can vary from 1, in standard monomodal dynamic AFM, to N>1 in the more recent multifrequency modes of operation[26, 28, 29]. In monomodal AFM, and particularly in ambient and vacuum environments, a single mode (m=1) is typically accounted for because the excitation of higher modes, and in general harmonics, is largely inhibited[10, 26]. Here, two external forces, i.e. N=2, have been included in (1) as in standard bimodal AFM as first introduced[26]. In bimodal AFM the second driving force externally introduces harmonic distortion in order to enhance the detection of higher harmonics[30], typically the one that is excited[31], at lower peak forces than monomodal AFM[32]. The effective external drive forces can be written in terms of experimental parameters as

$$F_{0m} = k_m A_{0m} \sqrt{\left(1 - \beta_m^2\right)^2 + \left(\frac{\beta_m}{Q_m}\right)^2} \qquad (2)$$

where $A_{0m}$ and $\beta_m$ are the modal free amplitude and the normalized drive frequency $\beta_m = \omega_m/\omega_{0m}$ respectively. When the tip interacts with the sample, the response of the cantilever in bimodal AFM can be written as

$$z(t) = z_0 + A_1 \cos(\omega_1 t - \phi_1) + A_2 \cos(\omega_2 t - \phi_2) + O(\varepsilon) \qquad (3)$$

where z(t) is the absolute tip position, i.e. a sum of $z_m(t)$ over m, $z_0$ is the mean deflection, $A_1$, $A_2$, $\phi_1$ and $\phi_2$ are the oscillation amplitudes and phase shifts at $\omega_1$ and $\omega_2$ respectively and $O(\varepsilon)$ stands for the higher harmonic contributions where the higher harmonics might be multiples of $\omega_1$, $\omega_2$ or both[30]. If $\omega_1 \approx \omega_{01}$ and $\omega_2 \approx \omega_{02}$ the response at the two relevant frequencies $\omega_1$ and $\omega_2$ can be reduced to the contributions from modes 1 and 2 only at these particular frequencies. In this work the resonance frequencies $\omega_{01}$ and $\omega_{02}$ coincide exactly with the drive frequencies $\omega_1$ and $\omega_2$ for simplicity unless otherwise stated and are integer multiples[33]. Note also that in this work the subscript for the first mode has been dropped when it does not lead to ambiguity.

A main objective of this work is to relate operational parameters to experimental observables and to discuss the relationship between these experimental observables and the tip-sample interaction in bimodal AFM with small amplitudes. Nevertheless, it is a typical condition of bimodal AFM that $A_2/A_1 \ll 1$ implying that cantilever dynamics will be mostly controlled by the fundamental mode. Thus, it is instructive to first discuss monomodal AFM in order to establish the operation regimes that can be accessed. Assuming that in monomodal AFM the response is well approximated by the fundamental frequency only, it follows that

$$z(t) \approx A_1 \cos(\omega_1 t - \phi_1) \tag{4}$$

Then,

$$d_m \approx z_c - A_1 \tag{5}$$

where $d_m$ is the minimum distance of approach and $z_c$ is the cantilever-sample separation. Controlling the distance $d_m$ is critical in dynamic AFM since this will determine peak forces[5, 11, 23], regime of operation[34] and, in general, lateral resolution[1], sample invasiveness and the sample properties that are probed in a given experiment. For the purpose of discussing the data in this work, the distance d is identified with $d_m$ and is used interchangeably throughout.

In Fig. 1a the behavior of $\bar{A}_1$ ($\bar{A}_1 = A_1/A_0$ and $A_0 \equiv A_{01}$) and $d_m$ have been monitored as a function cantilever separation $z_c$ while employing a free amplitude $A_0$ of 0.5 nm. The behavior of $d_m$ with $A_0$ and $\beta_1$ is discussed later. The cantilever parameters in this work (throughout) are: $k_1$, $k_2$ = 40 and 1600 N/m, $f_1$ = 300 kHz ($\omega_1 = 2\pi f_1$), $f_2$ = 1.8 MHz ($\omega_2 = 2\pi f_2$), $Q_1$ = 450 and $Q_2$ = 2700. Furthermore, the tip-sample force $F_{ts}$ has been modeled as follows;

1) In the long range, $F_{ts}$ is determined by the Hamaker constant H[22]

$$F_{ts}(d) = -\frac{RH}{6(d-h)^2}(\alpha_{nc}+1) \qquad\qquad h+a_0<d \qquad\qquad (6)$$

where d is the tip-sample distance, R is the tip radius, h is the height of the water layer[15] and $a_0$ is an intermolecular distance. In this work h=1 nm and $a_0$=0.165 nm throughout. The meaning of $\alpha_{nc}$ is discussed below.

2) When the tip is inside the water layer ($a_0<d<h+a_0$) $F_{ts}$ is constant and coincides with the adhesion force $F_{AD}$

$$F_{ts} \equiv F_{AD} = -\frac{RH}{6a_0^2}(\alpha_{nc}+1) \qquad\qquad a_0<d\leq h+a_0 \qquad\qquad (7)$$

3) When mechanical contact occurs ($d<a_0$) the Derjaguin Muller Toporov (DMT) model of contact mechanics assumes that the force is controlled by $F_{AD}$, the effective Young modulus in the contact $E^*$ and the tip-sample deformation $\delta$[35]

$$F_{DMT}(d) = F_{AD}(\alpha_c+1) + \frac{4}{3}E^*\sqrt{R}\delta^{3/2} \qquad\qquad a_0\leq d \qquad\qquad (8)$$

where the tip-sample distance d and $\delta$ are related by $\delta = a_0-d$. The meaning of $\alpha_c$ is discussed below.

4) Finally, dissipation has been assumed to occur in the form of hysteresis only. If no energy is dissipated in the tip sample interaction $\alpha_{nc}=\alpha_c=0$. The subscripts nc and c stand for noncontact and contact respectively. If energy is dissipated $\alpha_{nc}$ and/or $\alpha_c \neq 0$ on tip retraction.

In summary, the model described in points 1 to 4 above describes samples properties via hysteretic dissipation (chemistry and mechanics), the Hamaker constant H (chemistry) and the young modulus $E^*$ (mechanics in the contact region). Furthermore the force profile is consistent with force profiles in ambient conditions where a nanometer thick water film is typically present on the surface[34, 36, 37]. The last point is particularly relevant since it has been recently shown independently by Wastl et al.[3] and Santos et al.[11] that penetrating the water layer with small oscillation amplitudes might lead to higher lateral resolution and signal to noise ratio. More recently[22] enhanced contrast has been predicted by exciting the second flexural mode and penetrating the water layer as discussed in the above works.

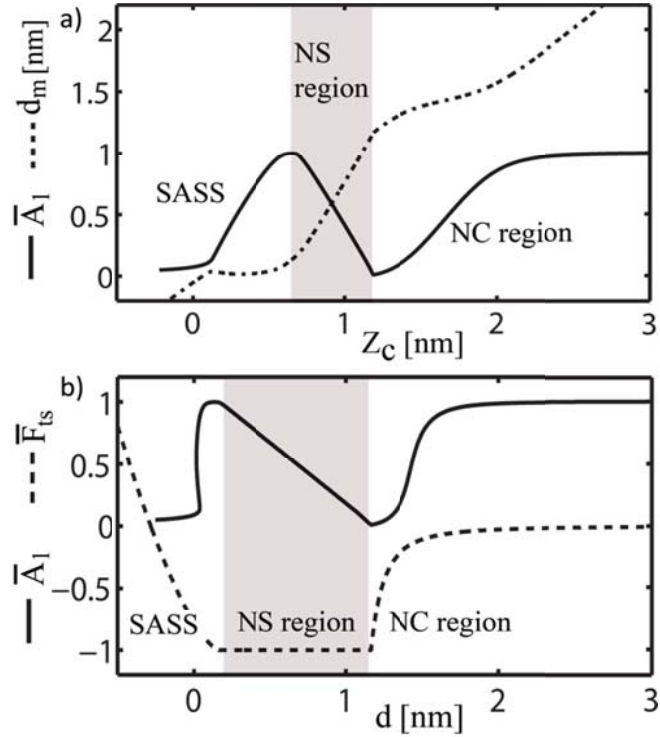

FIG. 1 (a) Response of the fundamental amplitude $\bar{A}_1$ (continuous lines) and the minimum distance of approach $d_m$ (dashed lines) as a function of separation $z_c$ with the use of small amplitudes $A_0$=0.5 nm. (b) Relationship between $\bar{A}_1$ and $d_m$ or $d$, i.e. $\bar{A}_1(d)$. The actual tip-sample (normalized) force $F_{ts}$ (dashed lines) is also plotted for comparison.

For the simulations in Fig. 1 $A_0$= 0.5 nm, $A_{02}$= 0 nm, $E_t$ =120 GPa (tip), $E_s$ = 1 GPa (sample), H=4.1×10$^{-20}$ J, R=3 nm, $\alpha_{nc}$=$\alpha_c$ =0 (no dissipation). The value R=3 nm has been employed throughout since high resolution is one of the themes of this work. The fact that $A_{02}$= 0 nm, i.e. nulled second external drive, implies that the second mode can only be excited by the tip-sample interaction. In Fig. 1a the amplitude response $\bar{A}_1$ (continuous black lines) and the minimum distance of approach $d_m$ (dashed lines in nm units) are shown as a function of separation $z_c$. There are two regions of positive slope in $\bar{A}_1$ where AM AFM can be operated and one of negative slope (NS region) which is not available for AM AFM operation. Such phenomena have been recently reported experimentally[11, 37] on several surfaces such as mica, graphite, aluminum and quartz[37]. The first positive slope coincides with the noncontact regime (NC region) of operation and the tip oscillates above the water layer $d_m > h + a_0$. Thus, physically, this region allows probing long range forces. In the NS region the tip makes contact with the water layer $a_0 < d_m < h + a_0$ and $d_m$ decreases sharply with $z_c$. Finally, for smaller separations, local maxima in $\bar{A}_1$ is observed and the slope is again inverted. This second region of positive slope was recently[11] termed small amplitude small set-point (SASS) and allows probing short range mechanical phenomena with small amplitudes. Furthermore in SASS the tip is typically in perpetual contact with the water layer[3, 11]. If $\bar{A}_1$ is further reduced, the separation $z_c$ decays much faster than $\bar{A}_1$ implying that deformation dramatically increases as it can be confirmed by the drastic change of slope in $d_m$ (Fig. 1a). The relationships between $d_m$, $\bar{A}_1$ and the tip-sample force $F_{ts}$ under the different conditions can be more clearly seen in Fig. 1b where $\bar{A}_1$ (continuous lines) and $\bar{F}_{ts}$ ($\bar{F}_{ts} = F_{ts}/|F_{AD}$, dashes lines) have been plotted against $d_m$ or d, i.e. $\bar{A}_1(d)$ and $\bar{F}_{ts}(d)$. Experimentally, $\bar{A}_1(d)$ can be approximately recovered via (5). In summary, the results presented in Fig. 1 imply that even when employing free amplitudes as small as 0.5 nm the full range of distances, down to the region of mechanical contact, can be probed. In practice however, noise and instabilities might be present when employing small amplitudes[7, 38, 39]. Furthermore, for a given experiment, the user might want to employ a given oscillation amplitude $\bar{A}_1$ while accessing a given minimum distance of approach $d_m$ or operation regime. Thus alternatives are discussed below in terms of variations of β and $A_0$.

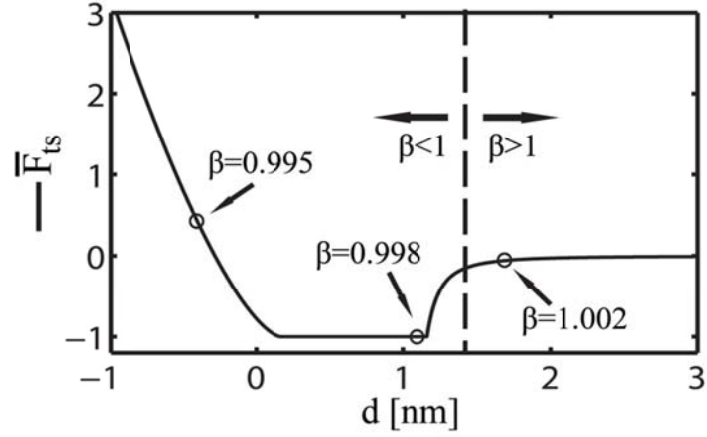

FIG. 2 Variation of fundamental drive frequency β under standard small amplitude operation $\bar{A}_1$=0.9. Numerical integration predicts that the minimum distance of approach can be varied for a given $\bar{A}_1$ (here 0.9) and $A_0$ (here 1 nm) as a function of fundamental (normalized) drive frequency β.

Conventionally, AM AFM users employ small free amplitudes $A_0$ and large set points $\bar{A}_1$≈0.8-0.9 to decrease peak forces and image in the NC region[6, 40]. In Fig. 2 the free amplitude and set point have been set to $A_0$= 1 nm and $\bar{A}_1$=0.9 while varying the drive frequency $β_1≡ β$. When driving exactly at resonance β=1 the minimum distance of approach d stays above h+$a_0$ (vertical dashed line) implying that the tip oscillates above the water layer as expected. Increasing β leads to larger values of d implying that the mean distance increases. See for example the value of d for β=1.002 (circle in Fig. 2). Nevertheless the behavior is non-monotonic and d might decrease for larges values of $β_1$ (not shown). Decreasing $β_1$ leads to a monotonic decrease in d (see circles in Fig. 2) even though the behavior is non-linear (not shown). In summary, one can potentially experimentally access any distance d with small amplitudes by varying β alone. It should be noted however that in Fig. 2 $A_0$ has been defined as the value of free amplitude at the given β. Physically this implies that $F_{01}$ has been modified to reach the given $A_0$ for a given β as predicted by (2).

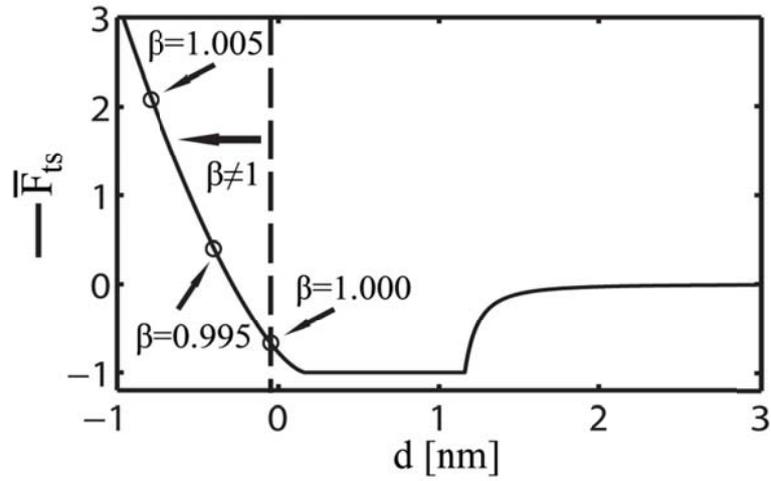

FIG. 3 Variation of fundamental drive frequency $\beta$ under SASS. Numerical integration predicts that the minimum distance of approach can be varied for a given $\bar{A}_1$ (here 0.2) and $A_0$ (here 1 nm) as a function of fundamental (normalized) drive frequency $\beta$ when the tip is in perpetual contact with the water layer (SASS).

The predictions of the variation in d with $\beta$ under the SASS conditions are shown in Fig. 3 where $A_0$= 1 nm and $\bar{A}_1$=0.2. In this case d monotonically decreases for both increasing and decreasing $\beta$ (see circles for specific values) implying that minima in tip-sample deformation should occur exactly at $\beta$=1. Nevertheless, again, the behavior is non-linear (not shown).

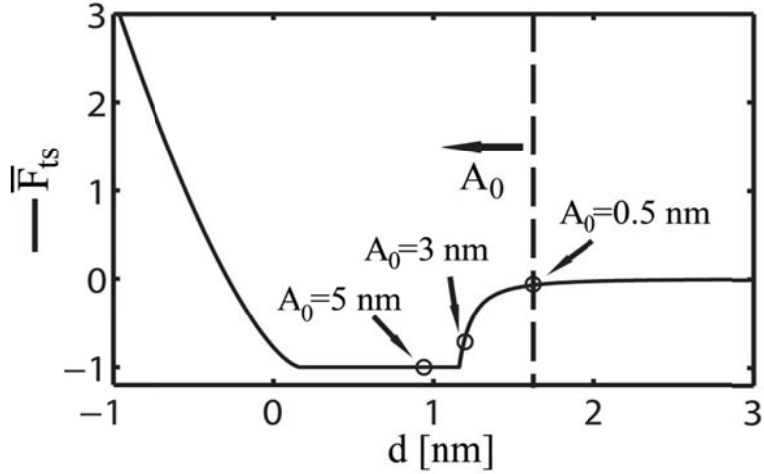

FIG. 4 Variation of fundamental free amplitude under standard small amplitude operation $\bar{A}_1=0.9$. Numerical integration predicts that the minimum distance of approach can be varied for a given $\bar{A}_1$ (here 0.9) as a function of $A_0$ when driving exactly at $\beta=1$.

One may also vary the free amplitude $A_0$ alone while staying exactly at $\beta=1$. For example, in Fig. 4 $A_0$ has been increased from 0.5 nm to 5 nm while keeping the set point at $\bar{A}_1=0.9$. In this case, the minimum distance of approach d monotonically decreases with increasing $A_0$ until the water layer is eventually contacted for sufficiently high values of $A_0$ (see circle for $A_0=$ 5 nm in Fig. 4). In SASS the set-point is typically set to 0.5 nm or less (see Fig. 1) by default[3, 11]. Thus in Fig. 5, $A_0$ has been increased from 0.5 nm to 5 nm while keeping the set-point at $A_1=0.2$ nm. The result is a monotonic and non-linear decrease in d with increasing $A_0$. In summary, while minimal invasiveness results for the largest values of d in order to increase resolution and probe mechanical properties d needs to be sufficiently small [5, 6, 9, 41]. More generally, the full range of sample properties, i.e. short range and long range, can be probed by controlling d. In this respect, the above results show that small amplitudes can lead to carefully and controllably operating the AFM in the full range of distances of interest by increasing and decreasing d via the standard operational parameters $\beta$ and $A_0$.

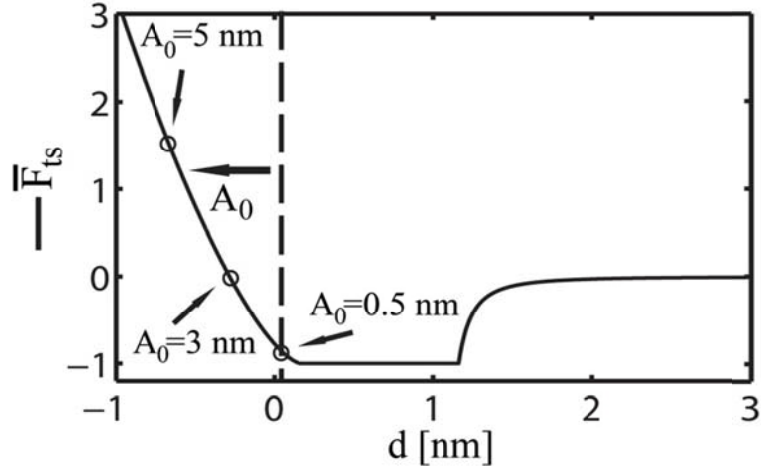

FIG. 5 Variation of fundamental free amplitude under SASS. Numerical integration predicts that the mechanical contact region can be accessed and that the minimum distance of approach d can be decreased for a given $A_1$ (here 0.2 nm) with decreasing $A_0$ in SASS. Here $\beta=1$.

### III. Enhancing contrast in bimodal AFM with small amplitudes

According to section II, the minimum distance of approach can be controlled by the user in AM AFM in a given experiment by manipulating $A_0$ and/or $\beta$. This provides the means to explore a given set of sample properties and/or experimentally accessing the different force regimes even when employing small amplitudes. Next, the second flexural mode is externally excited ($F_{02}>0$ in (1)) in order to open two extra channels, i.e. $A_2$ and $\phi_2$, that provide access and/or enhance compositional contrast.

First, multiplying (1) by (3) and integrating over a fundamental period T gives

$$V_m \equiv \frac{1}{T}\oint F_{ts} z_m dt = -\frac{1}{2} F_{0m} A_m \cos\phi_m \tag{9}$$

where $V_m$ stands for virial. In bimodal AFM the suffix m can be 1 or 2 standing for first and second flexural modes respectively. The net tip-sample virial is the sum of these

$$V_{ts} = V_1 + V_2 \tag{10}$$

If (1) is multiplied by the time derivative of (3) the energy transfer (transferred between modes or onto the tip-sample junction) is obtained

$$E_1 \approx \frac{\pi k_1 A_0 A_1}{Q_1}\left[\sin\phi_1 - \frac{A_1}{A_0}\right] \tag{11}$$

$$E_2 \approx \frac{n\pi k_2 A_{02} A_2}{Q_2}\left[\sin\phi_2 - \frac{A_2}{A_{02}}\right] \tag{12}$$

where $E_1$ and $E_2$ are the energy transferred via modes 1 and 2 respectively and $n=\omega_2/\omega_1$ (in the simulations here $n=6$ and $\omega_1/\omega_{01}=\omega_2/\omega_{02}=1$). The main approximation in (11) and (12) is that harmonics other than the fundamental frequencies of each mode have been ignored. The energy dissipated in the tip-sample interaction $E_{dis}$ is

$$E_{dis} = E_1 + E_2 \tag{13}$$

implying that $E_2$ can also be written as $E_{dis}-E_1$. Furthermore the modal phase shifts are

$$\sin\phi_1 \approx \frac{A_1}{A_0} + \frac{Q_1}{\pi k_1 A_0 A_1} E_1 \tag{14}$$

$$\sin\phi_2 \approx \frac{A_2}{A_{02}} + \frac{Q_2}{n\pi k_2 A_{02} A_2} E_2 \tag{15}$$

Recently, the relative kinetic energy between modes has been investigated and related to the emergence of three modes of operation in multifrequency AFM[31, 42]. Here however (13) is discussed in terms of energy transfer between modes and energy dissipation $E_{dis}$. If the interaction is fully conservative $E_{dis}=0$ and (13) implies that the energy transfer between modes is $E_1=-E_2$.

If the transfer of energy between modes is zero or small, $E_1 = -E_2 \approx 0$ and, provided $E_{dis}=0$, a further approximation follows for the phase shifts

$$\sin\phi_m \approx \frac{A_m}{A_{0m}} \qquad (16)$$

in agreement with previous studies [30, 43]. (16) does not provide information about the source of contrast in second mode shift $\phi_2$. Information however can be extracted from the virials $V_m$ in Eq. (9). In particular $V_1$ is known[38] to relate to conservative interactions and the conservative contribution to $F_{ts}$ can be extracted from it [44, 45]. On the other hand, $V_2$ has been related to the frequency shift of the second flexural mode $\Delta\omega_2$ [43, 46]

$$V_2 \approx -A_2^2 k_2 \frac{\Delta\omega_2}{\omega_{02}} \qquad (17)$$

Furthermore, provided $A_2 \ll A_1$, the time averaged derivative of the tip sample force $<F_{ts}'>$ per fundamental oscillation period T can be written as[47]

$$<F_{ts}'> \equiv \frac{1}{T}\oint \frac{dF_{ts}}{dt} dt \approx -2k_2 \frac{\Delta\omega_2}{\omega_{02}} \qquad (18)$$

Then, from (9), (17) and (18) it follows that

$$<F_{ts}'> \approx -\frac{F_{02}}{A_2}\cos\phi_2 \qquad (19)$$

where all the parameters on the right hand side of (19) are experimental observables in AM AFM. We note that the above equations could, in principle, be employed in both AM and frequency modulation (FM) AFM. Furthermore, while Eq. (19) implies that the phase shift of the second mode $\phi_2$ is related to conservative interactions only, i.e. $<F_{ts}'>$, care should be taken since dissipation alone might lead to variations in the minimum distance of approach d[5, 48] and thus in the value of $<F_{ts}'>$. This will be shown later. Importantly, neither energy transfer between modes nor energy dissipation appears in (19). That is, (19) indicates that situations of zero energy transfer between modes, i.e. $E_1=E_2=0$ in (14) and (15), and zero energy dissipation, i.e. $E_{dis}=0$ in (13), can still lead to second mode phase contrast provided there are variations in $<F_{ts}'>$. This is consistent with (16). Furthermore, from (16) and (19)

$$<F_{ts}'> \approx \pm \frac{k_2}{Q_2}\sqrt{\left(\frac{A_{02}}{A_2}\right)^2 - 1} \qquad (20)$$

where the term inside the square root will never be negative since energy transfer between modes is assumed to be zero in the derivation. That is, if $A_2>A_{02}$, energy has been transferred between modes and (16) does not stand. An expression for $\phi_2$ can also be written in terms of $V_2$ and $E_2$[30]. Note however that here $E_2$ should be considered as a combination of energy transfer between modes and energy dissipation and not energy dissipation alone. The expression is

$$\tan\phi_2 \approx \frac{-1}{2V_2}\left(\frac{E_2}{n\pi} + \frac{k_2 A_2^2}{Q_2}\right) \qquad (21)$$

or more restrictively ($A_2 \ll A_1$)

$$\tan\phi_2 \approx \frac{-1}{<F_{ts}'>}\left(\frac{E_2}{n\pi A_2^2} + \frac{k_2}{Q_2}\right) \qquad (22)$$

The sources of contrast via $A_2$ and $\phi_2$ can thus be divided into three cases:

1) No irreversible loss of energy, i.e. $E_{dis}=0$ in (13), and no energy transfer between modes, i.e. $E_n=0$ in (11) and (12), occur. In this case, the source of contrast has to be traced back to the Virial of the second mode $V_2$ as expressed by (9). Since in bimodal

AFM the second mode is typically left open loop, variations in $V_2$ will translate into variations in $A_2$ and $\phi_2$ according to (9), (16) or (19). The physical source of contrast in this case is arguably made more intuitive by looking at (22) and assuming $E_2=0$. Computationally $\phi_2$ can be trivially obtained from (16). In summary variations in $\phi_2$ originate from variations in $<F_{ts}'>$.

2) No irreversible loss of energy, i.e. $E_{dis}=0$ in (13), but significant energy transfer between modes, i.e. $E_n \neq 0$ in (11) and (12), occur. In this case variations in $\phi_2$ originate from variations in $V_2$ or $<F_{ts}'>$, as expressed in (9) or (19), and variations in energy transfer between modes $E_2$ as in (12). These are all conservative interactions and the contributions are accounted for in (22) in both the terms inside and outside the brackets. Computationally, these contributions are accounted for in (15) by the first and second terms on the right respectively.

3) Irreversible loss of energy, i.e. $E_{dis} \neq 0$ in (13), and significant energy transfer between modes, i.e. $E_n \neq 0$ in (11) and (12), occur. The origin of contrast in this case is similar to that in case 2 above but now $E_2$ has contributions from dissipative interaction s $E_{dis}$ and energy transfer between modes.

Equations (11) to (22) can be compared to simulations to test the theory. In particular, in Fig. 6 the interaction is conservative and $\alpha_{nc}=\alpha_c=0$. In Fig. 7 there is dissipation and $\alpha_{nc}=10\alpha_c=0.1$. Numerical integration has been carried out employing the Euler method and the (fourth order) Runge Kutta algorithm. Conservation of energy followed from the former but not from the latter implying that for accurate calculations of energy dissipation and energy transfer the Euler method is preferred. In particular, by allowing $2^{12}$ points per cycle, i.e. step time ≈0.4 ns, with the Euler method errors of 0.1 meV per cycle followed and these were reduced to 1 peV by allowing $2^{14}$ points per cycle. The error with the Runge Kutta algorithm could not be reduced below 0.1meV for any step time.

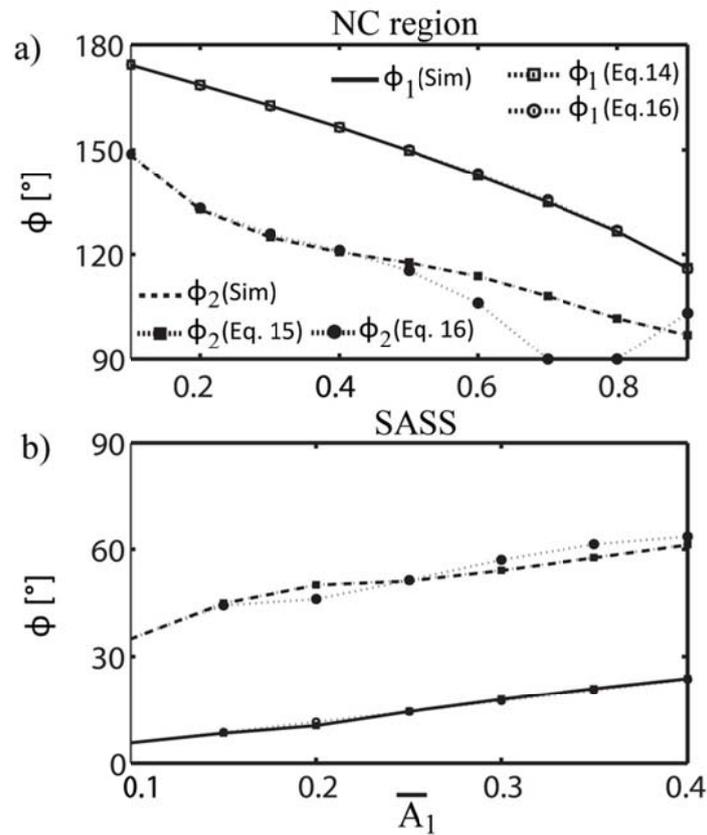

FIG. 6 Numerically calculated (Sim) response of the fundamental shift $\phi_1$ (continuous lines) and the second mode shift $\phi_2$ (dashed lines) as a function of normalized amplitude $\bar{A}_1$ in the (a) NC region and the (b) SASS region. Comparisons with analytical expressions are shown. The interaction is conservative.

Fig. 6 shows the results of calculating the phase shift of the first $\phi_1$ (continuous lines) and second $\phi_2$ (dashed lines) modes in the simulations (Sim) when employing $A_0$=1 nm and $A_{02}$=50 pm (0.1<$\bar{A}_1$<0.9). Fig. 6a shows the results obtained when the tip oscillates in the NC region (d>h+$a_0$) as defined when discussing Fig. 1. Fig. 6b shows the results obtained in the SASS region also as discussed in Fig. 1. For the SASS region the range of oscillation amplitudes is 0.1<$\bar{A}_1$<0.4 and the tip always oscillated in perpetual contact with the water layer. Eqs. (14) and (15) match the phase shift in the simulations (Sim) with errors smaller than 0.1°. On the other hand fractions of a degree and even several degrees of error in phase shift can follow when employing the approximation in Eq. (16), particularly in the second mode shift $\phi_2$. The implication is the in some cases energy transfer between modes cannot be ignored when interpreting $\phi_2$. In Fig. 7, $E_{dis}$>0 and the approximations given by (14) and (15) are shown together with the results of the simulations (Sim). Again, errors smaller than 0.1° followed from (14) and (15) throughout both in the NC region (Fig. 7a) and in the SASS region (Fig. 7b).

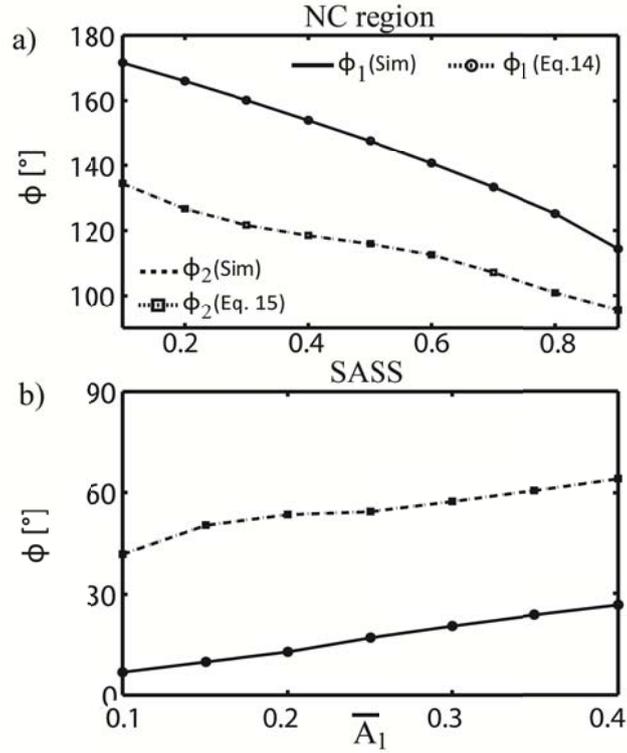

FIG. 7 Numerically calculated (Sim) response of the fundamental shift $\phi_1$ (continuous lines) and the second mode shift $\phi_2$ (dashed lines) as a function of normalized amplitude $\bar{A}_1$ in the (a) NC region and the (b) SASS region. Comparisons with analytical expressions are shown. The interaction includes dissipation.

It remains to be seen however how $\phi_1$ and $\phi_2$ respond to variations in conservative and dissipative sample properties in the NC and SASS regions with small amplitudes. This issue can be addressed by allowing variations in the simulations and comparing the phase shifts and $E_{dis}$ (dissipative) and $\langle F_{ts}' \rangle$ (conservative). In particular, one can define a difference in phase shift for two sets of sample parameters $\Delta\phi_m = \phi_m(\text{sample 2}) - \phi_m(\text{sample 1})$ where sample 1 and 2 stand for two surface locations presenting variations in sample properties.

In Fig. 8a the Hamaker constant has been varied from $H=4.1\times10^{-20}$ J (sample 1) to $6.1\times10^{-20}$ J (sample 2) leading to a variation of $\Delta H=2\times10^{-20}$ J. The free amplitudes were $A_0=1$ nm and $A_{02}=50$ pm. In the figures, differences in phase shift $\Delta\phi_m$ have been normalized relative to maxima. The results of the numerical integration of (1) in Fig. 8a show that $\Delta\phi_2$ (squares and dashed lines) is typically about an order of magnitude, or more, larger than $\Delta\phi_1$ (circles and continuous lines). This implies higher sensitivity of $\Delta\phi_2$ to variations in conservative interactions. In particular, maxima of ≈3.3° degrees resulted in $\Delta\phi_2$ for $\bar{A}_1\approx0.5$ (Fig. 9a). The fact that $\Delta\phi_1\neq0$ however implies that the first mode shift is also sensitive to conservative interactions. This is a result of energy transfer similarly as recently discussed in liquid[27] and as dictated by (14) when $E_1\neq0$. In particular $\Delta\phi_1\neq0$ resulted in Fig. 8 from energy transfer no larger than 10 meV. It is also worth noting that the variation in Hamaker employed in Fig. 9a, i.e. $\Delta H=2\times10^{-20}$ J, is similar to that between a silicon tip and a mica surface and a silicon tip and silicon nitride surface[49]. The implication is that chemical variations in these systems should be detected with phase shifts as large as 3° with cantilever specifications such as those employed here, i.e. similar to those of AC160TS (Olympus). In the SASS region (Fig. 8b) the Young modulus of the sample $E_s$ has been varied from 1 (sample 1) to 5 GPa (sample 2) while maintaining all other parameters as above. Such variations in $E_s$ should be typical of polymers. Again, the second phase shift is more sensitive since maxima in Fig. 8b is $\Delta\phi_1\approx1.9°$ and $\Delta\phi_2\approx10.2°$. Maxima in $<F_{ts}'>$ was $<F_{ts}'>\approx0.035$ [N/m] in the NC region and ≈0.129 [N/m] in the SASS region. In summary, Fig. 8 exemplifies two important aspects of bimodal AFM. First, while both phase shifts might provide information about conservative compositional variations, the shift of the second mode can be an order of magnitude larger, or more, for a given variation in sample properties. Second, variations in both long (NC) and short (SASS) range material composition can be probed with the use of small amplitudes via $\phi_2$ even where no energy dissipation occurs. This is in agreement with the predictions in (9), (16), (19) and (22).

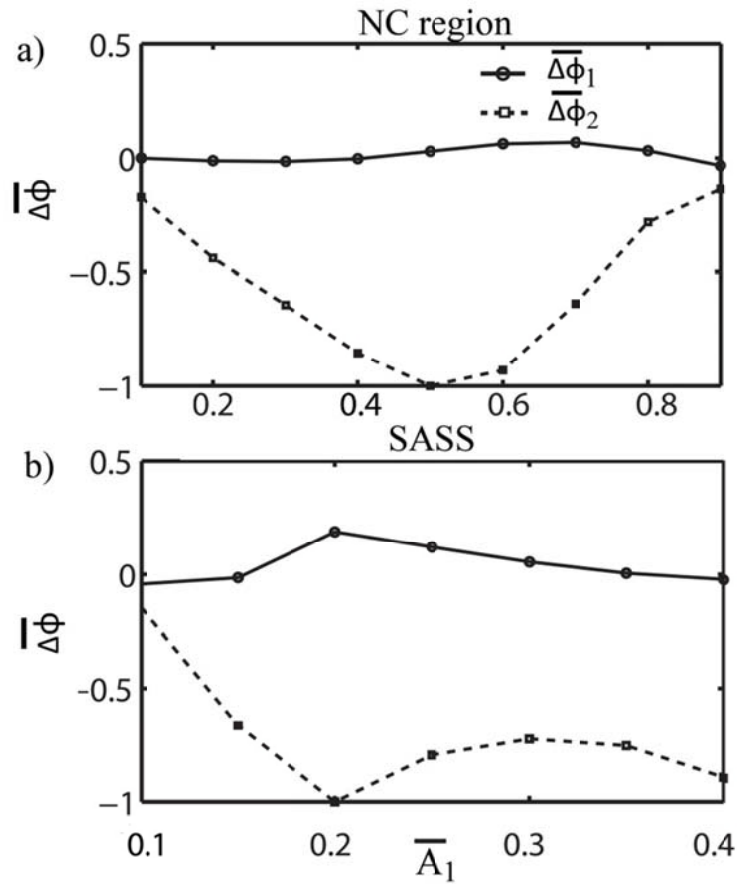

FIG. 8 Numerically calculated difference (normalized) in phase shift response of the first $\Delta\phi_1$ (continuous lines) and second $\Delta\phi_2$ modes (dashed lines) as a function of normalized amplitude $\bar{A}_1$. Conservative variations only are allowed in the interaction in (a) the NC (maxima of $\Delta\phi_2 \approx 3.3°$) and (b) SASS (maxima of $\Delta\phi_2 \approx 10.2°$) regions.

The behavior of $\Delta\phi_m$ when variations in dissipative properties of the material only are present is discussed with the use of Fig. 9. In particular, here the parameters are the same as those in Fig. 1 and $A_0$=1 nm, $A_{02}$=50 pm, $\alpha_{nc}=\alpha_c=0$ (sample 1) and $\alpha_{nc}=10\alpha_c=0.1$ (sample 2). Now the first and second differences in phase shifts $\Delta\phi_2$ and $\Delta\phi_1$ are of the same order of magnitude except at the smaller set points, i.e. $\bar{A}_1\ll 1$, where $\Delta\phi_2\gg\Delta\phi_1$. Maxima was $\Delta\phi_2\approx 14.2°$ in the NC region (Fig. 9a) and $\approx 6.9°$ in the SASS region (Fig. 9b). The corresponding energy dissipated was 35 meV and 42meV respectively implying that energies in the order of 1 single van der Waals bond can lead to second mode shifts an order of magnitude larger than the noise level, i.e. $\approx 0.1$-$0.2°$. The fact that both modes are sensitive to variations in energy dissipation is consistent with the literature[43] and with the theory above. In particular the results are in agreement with (11)-(15).

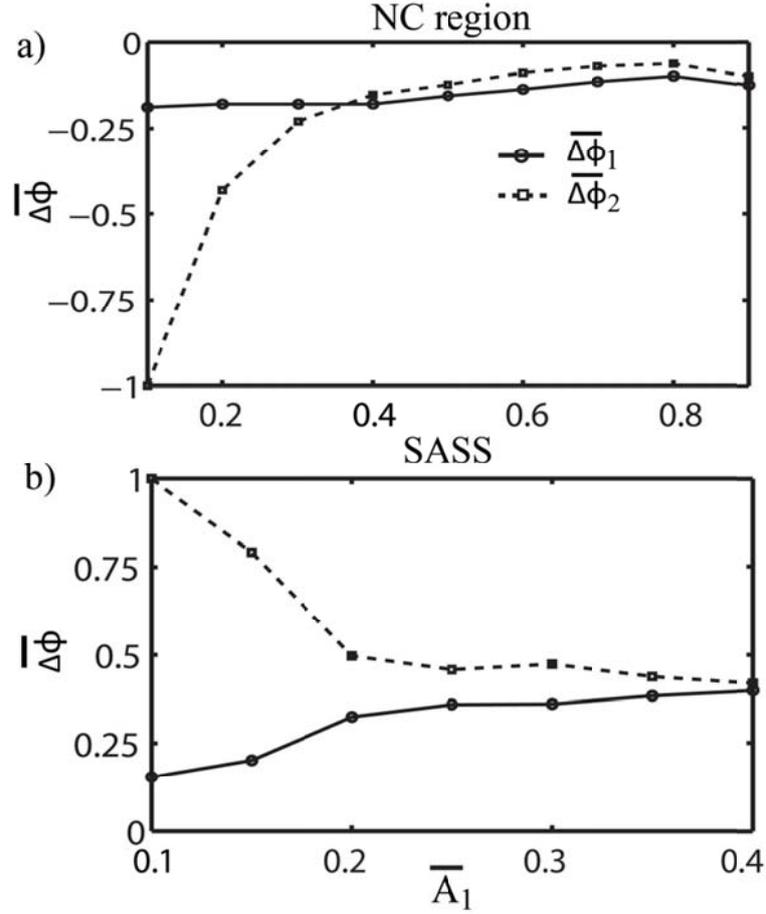

FIG. 9 Numerically calculated difference (normalized) in phase shift response of the first $\Delta\phi_1$ (continuous lines) and second $\Delta\phi_2$ modes (dashed lines) as a function of normalized amplitude $\bar{A}_1$. Dissipative variations only are allowed in the interaction in (a) the NC (maxima of $\Delta\phi_2 \approx 14.2°$) and (b) SASS (maxima of $\Delta\phi_2 \approx 6.9°$) regions.

Finally, a definition is given here in order to quantify the sensitivity of the phase shift difference of the second mode $\Delta\phi_2$ to variations in dissipative and conservative properties. For conservative interactions the sensitivity $S_{con}$ can be defined in terms of degrees per unit of stiffness [°/N/m]

$$S_{con} = \frac{\Delta\phi_2}{\Delta <F_{ts}'>} \tag{23}$$

For dissipative interactions the sensitivity $S_{dis}$ is defined in terms of degree per eV [°/eV] as

$$S_{dis} = \frac{\Delta \phi_2}{\Delta E_{ts}} \quad (24)$$

The parameters from Fig. 8 have been employed to characterize $S_{con}$ for a system presenting variations in conservative properties only, i.e. $\Delta H = 2 \times 10^{-20}$ J and $\Delta E_s = 4$ GPa, in the NC region (squares and continuous lines) and in the SASS region (circles and dashed lines) respectively. The results are presented in Fig. 10 where the sensitivity $S_{con}$ has been normalized with maxima giving $S_{con} \approx 99.7$ [°/N/m] at approximately $\bar{A}_1 = 0.6$ in the NC region. Maxima of $\Delta <F_{ts}'>$ of $\approx 0.13$ N/m resulted in the SASS region. It is physically and practically relevant to note that maxima in contrast $\Delta \phi_2$ (see Fig. 8a) does not coincide with maxima in sensitivity $S_{con}$, i.e. in the NC region maxima occurred at $\bar{A}_1 \approx 0.5$ for $\Delta \phi_2$ versus $\bar{A}_1 \approx 0.6$ for $S_{con}$. In general, the behavior of $S_{con}$ in both the NC and the SASS regions is nonmonotonic with decreasing $\bar{A}_1$.

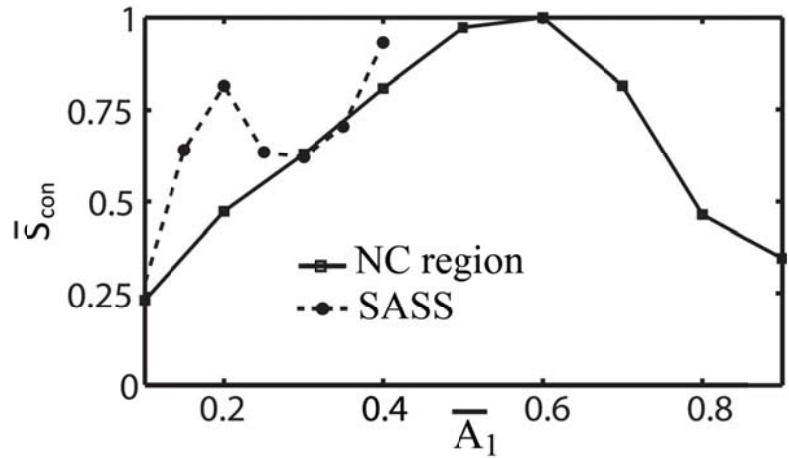

FIG. 10 Numerically calculated sensitivity (normalized) $S_{con}$ of the second mode phase shift $\Delta \phi_2$ to variations in conservative sample properties in the NC (continuous lines) and SASS (dashed lines) regions. The parameters as the same as those in Fig. 8. Maxima was $S_{con} \approx 99.7$ [°/N/m].

In Fig. 11 the dissipative parameters only have been varied as in Fig. 9, i.e. $\alpha_{nc} = \alpha_c = 0$ (sample 1) and $\alpha_{nc} = 10\alpha_c = 0.1$ (sample 2). This resulted in variations in both $S_{con}$ and $S_{dis}$. Physically, this implies that variations in dissipation only induce variations in the minimum distance of approach[5] and these affect $<F_{ts}'>$. In particular, maxima in $\Delta<F_{ts}'>$ resulted in the SASS region with ≈0.15 N/m and maxima in $\Delta E_{dis}$ also resulted in the SASS region with 42meV. Maxima in $S_{con}$ and $S_{dis}$ was 1066 [°/N/m] and 1092 [°/eV] respectively. For convenience absolute values have been plotted in Fig. 11 and normalized numerically with 1092.

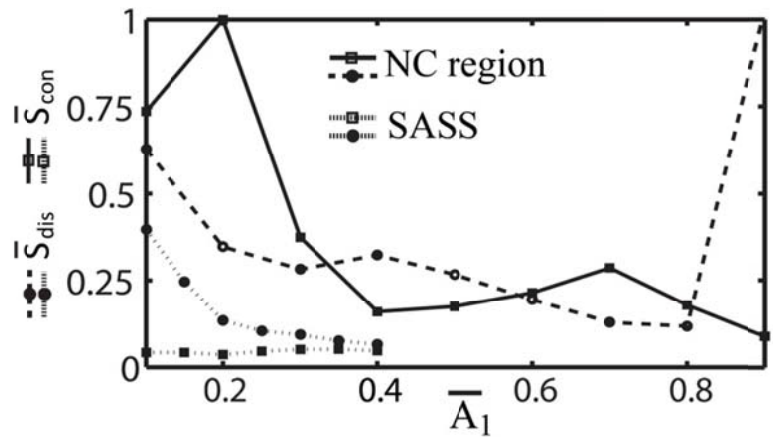

FIG. 11 Numerically calculated sensitivity (normalized) $S_{dis}$ in the NC (dashed lines and circle markers) and SASS (thin dashed lines and circle markers) regions in the presence of dissipative variations in sample properties only. The signal $S_{con}$ has also been plotted with continuous lines and square markers and thin dashed lines and square markers in the NC and SASS regions respectively. Maxima was $S_{dis}$≈1092 [°/eV].

IV. Conclusions

In summary, the theory of bimodal AFM operated with amplitudes comparable to intermolecular bonds or small molecules has been discussed in the context of ambient AFM. It has been show that the tip can be made to oscillate in a controllable fashion, and with the use of standard parameters, in a range of distances of interest for high resolution and minimally invasive mapping. These include the non-contact region where the tip oscillates above the hydration layer and the region where the tip mechanically interacts with the surface under the hydration layer. The tip-sample distance can be controlled in these operation regimes by varying the standard operational parameters.

By exciting the second mode with sub-angstrom amplitudes, the second mode phase shift becomes readily accessible for mapping compositional contrast with enhanced sensitivity. In the presence of variations in conservative sample properties only, the first and second mode phase shifts provide compositional contrast via variations in conservative interactions and transfer of energy between modes. The second phase shift however is typically an order of magnitude larger for a given interaction. In the presence of variations in dissipative sample properties only, compositional contrast is provided via variations in tip-sample distance induced by dissipation, the transfer of energy between modes and irreversible loss of energy in the tip-sample junction.